# Physics as a Mechanism for Including ELLs in Classroom Discourse

Enrique Suarez and Valerie Otero

*School of Education, University of Colorado, Boulder, 80309, USA*

**Abstract.** English Language Learners (ELLs) are frequently left on the periphery of classroom interactions. Due to misalignment of language skills, teachers and peers communicate with these students less often, decreasing the number of opportunities to engage. Exclusion can be avoided with learning activities that invite all students to participate and contribute ideas. We argue that environments and activities that privilege scientific inductive reasoning increase possibilities for emerging bilingual students to engage. This study investigated first-grade students' discussions about factors that affect how objects float. Students came from a variety of language backgrounds; all were considered beginner/intermediate ELLs. Results show that the goal of inducing principles from actual phenomena encouraged students to communicate their ideas and reasoning, boosting students' confidence in expressing themselves. Following the hybrid space argument of Vygotsky's theory of concept formation, we illustrate that physics can be particularly suitable context for the co-development of concepts and English language skills.

**Keywords:** English Language Learners; elementary school; student engagement; scientific induction; buoyancy.
**PACS:** 01.40.eg, 01.40.Fk, 01.40.Ha

## INTRODUCTION

With the expected shifts in our country's population, there will be an increase in the number of emerging bilingual students in schools. Already a large number of students in the educational system are learning English as a second language, with a small number of teachers trained to meet their language needs [1]. Because of marked difference in language skills between teachers and English Language Learners (ELLs), as well as between monolingual and bilingual students, ELLs tend to not be fully incorporated into classroom activities. While English as a Second Language (ESL) classes are important, the development of language skills should also be part of other school subjects.

Because of pressures to increase language fluency of ELLs, there is a nationwide trend of teachers and administrators to decrease the amount of classroom time devoted to science [2]. However, reducing the amount of time spent on teaching and learning science may be counterproductive for students' language needs. We argue that a science classroom centered on evidence-based and inductive reasoning is particularly well-suited for engaging emerging bilingual students. By capitalizing on students' natural curiosity and desire to understand phenomena, these spaces push the boundaries of students' conceptual understanding. Additionally, when students participate in the process of scientific induction, we expect them to engage with peers in discussions about tangible, shared observations and questions that are meaningful.

While there have been efforts to understand the role language plays when ELLs learn science [3,4], there is still much research needed to explore how to use physics to increase engagement of these students in classroom activities. This is especially important since vocabulary and discourse practices can influence students' participation in inquiry and argumentation. Should the linguistic complexity of the environment be too high, these spaces will most likely exclude emerging bilingual students from activities. With this consideration, we set out to explore the following research questions: (i) what features of a classroom that engages ELLs in scientific induction foster student engagement?; and (ii) how can this engagement support students' conceptual and linguistic development? To answer these questions, we analyze how a group of 1$^{st}$ grade emerging bilingual students discussed the characteristics of boat's ability to float.

## CONCEPTUAL FRAMEWORK

We view learning as a social practice [5,6] and assume that both language skills and science concepts can (and should) be co-constructed in contexts that allow students to try out ideas, words, and identities in the presence of others. For emerging bilingual students, these communication expectations involve a different set of challenges than for monolingual English speakers. Specifically, students' language skills and discourse practices tend to be very different than the ones used and valued in academic settings. This incongruence often leads to the exclusion of some students from classroom interactions, which significantly hinders learning [4].

We hypothesize that for ELLs, the process of scientific induction is an area in which physics has a clear advantage over other disciplines. This is because students have the opportunity to practice evidence-based reasoning and generalizations of claims, all while participating in a public, collaborative discussion about shared and tangible experience. By partaking in these kinds of activities, we claim students learn.

Researchers have proposed Productive Disciplinary Engagement [7] as a construct for describing active student participation. Specifically, this type of engagement is defined as one in which students spontaneously participate, substantially contribute, and attend to each others' ideas, in a way that resembles disciplinary discourse practices and furthers intellectual progress. Four measures have been suggested to evaluate whether a learning environment can foster this type of engagement. First, teachers should encourage students to problematize the content through questions, proposals, and challenges. Second, it is important for "students to be authors and producers of knowledge (…) rather than mere consumers" [7; p. 404]. Third, students should be held accountable, particularly by how their work is responsive to what insiders and outsiders have established. And finally, it is necessary for students to be provided intellectual and material resources that can aide the sense-making process [7].

We see almost a one-to-one correspondence between the four principles discussed above and what we would expect to find in a learning environment based on inductive reasoning. To make the connection apparent, here we clarify what we mean by scientific induction. The first step of the process consists in observing physical phenomena and collecting evidence, followed by testing and postulating claims about mechanisms supported by the data. Subsequently, these evidence-based claims are communicated to the scientific community at large, through a peer review process, in which the validity and explanatory power of the principles is tested. Whether the community arrives at a consensus determines the creation of principles that are included in the larger corpus of knowledge, or if the original claims need revisions.

Classrooms that promote the inductive process almost effortlessly meet the four criteria for productive disciplinary engagement: students problematize content through their observations and are authors of the evidence-based claims; and these abstractions of are submitted to, and evaluated by, other students in the community. Interestingly, physicists and educators from the early 1900s embraced inductive reasoning as the appropriate method for teaching physics in high schools. In particular, C. R. Mann [8] commented that experiments should precede the statement of principles and laws. In his opinion, anchoring scientific principles on lived experiences is critical for student conceptual understanding.

This way of thinking resembles Vygotsky's Theory of Concept Formation [9], which proposes that learners inhabit informal spaces, populated by everyday experiences and interactions with the physical world, and academic spaces that house formal principles made available and validated by schools. During the process of concept formation learners develop conceptual understanding through grounding academic concepts in everyday experiences, while leveraging these particular instances for drawing generalizable, academic concepts [10].

While scientific induction could foster productive disciplinary engagement for ELLs, there is a big assumption with regards to the level of language dexterity required for students to participate. Therefore, it is important to consider what features of the learning environment will likely include ELLs and address their linguistic needs. In a sense, what is needed is the creation of third spaces [5,11] where students can recruit everyday language and discourse customs when sharing ideas associated with formal terminology and classroom practices.

There are scores of models for how individuals acquire a second language and, eventually, become bilingual. Krashen's Monitor Model has been appealing to educators because of its clear design guidelines for fostering language acquisition. Krashen criticizes drill-based, vocabulary-and-rules-first forms of language instruction, the type that is typically found in formal learning environments. In contrast to the decontextualized nature of this process, Krashen proposes that language should be developed through an unconscious process that focuses on language *as a means for engaging in relevant tasks*, rather than as a decontextualized goal itself [6].

Favoring the development, of a language through engagement, Krashen recommends learning environments that create opportunities for learners to partake in authentic discussions about an experience that is relevant to students. He warns that the complexity of the language used should always be accessible to every participant, otherwise students' communicative confidence may be negatively affected. Also, since the goals are to foster learners' confidence and the development of language skills through solving tasks, corrections on use of language should be avoided.

Taking all this into account, our hypothesis is that when ELLs are the authors and evaluators of evidence-based claims generated by shared, tangible experiences, they will experience productive disciplinary engagement and simultaneously further their conceptual understanding and language skills. To address this hypothesis we ask: (i) what features of a classroom that engages ELLs in scientific induction foster student engagement?; and (ii) how can this engagement support students' conceptual and linguistic development?

## RESEARCH CONTEXT

Twenty-one first grade students participated in this study. They were enrolled in a large K-8 urban public school that ran two separate academic programs: "Mainstream" for monolingual students and students proficient in English; and Sheltered English Immersion Program (SEIP) for students who were deemed to have limited proficiency. The school's demographic composition is as follows: 66% of students were ELLs; 76% qualified for Free and Reduced Lunch; 45% of students were Hispanic, 31% White, 13% Asian, and 9% African American.

Data were collected from a beginner/intermediate SEIP classroom. Seven different first languages were spoken, and students and their families came from ten different countries. Students' length of residence in the country ranged from US-born to arriving up to eight months before the session we present below.

The episode reported here was part of the unit on Engineering Design. The primary goals of this unit were for students to develop an appreciation for iterative design and to further their understanding of the physical variables that affected their designs. The unit was developed by the teacher, Krysta (pseudonym), and the first author, and consisted of students reading a short fictional story and agreeing on a problem to solve. The class chose to create anything that would help a mouse cross a deep river, which resulted in groups building boats or bridges. Each group drew a model, built the object, tested the object and indentified areas of improvement, bettered the prototype, and tested it one last time, each stage happening in 75-minute sessions.

## METHODOLOGY

We videotaped each of the five sessions from the Engineering Design unit. For this study, we analyze the third session, in which students tested projects they had designed and built in the two previous sessions. The class used a small plastic toy mouse as a stand-in for the story's character and filled a large vat with water to simulate the river. For testing, each group placed their project in the container of water and put the mouse on the boat or bridge. Diana and Sarita's (pseudonyms) boat had a circular piece of floral foam as its base, which was attached to the boat with a square dish sponge. The boat had a cardboard mast with feathers on top, and two plastic spoons coming off the sides of the floral foam base. Diana and Sarita's boat led to substantive discussion among students and is particularly useful for pointing out features of the learning environment that support student's engagement. This discussion is analyzed below. Data were in the form of video that was coded according to themes that emerged from the literature and data.

## FINDINGS

We analyzed conversations regarding Diana and Saritas's boat, specifically on students' predictions and discussion of testing results. Before testing began, the teacher asked students to predict what would happen when placing the boat in the river. Students expected the two spoons would make the boat more stable by preventing sideways motion. After predictions, Diana gently placed the boat on the river, which immediately tipped forward and threw the mouse into the water. They argued that the square dish sponge, rather than the circular foam, should have been at the bottom, "because squares float better than circles."

The most elaborate contribution came from Ellie (pseudonym), a Chinese student who had arrived in the US at the beginning of the school year. She had developed basic language skills throughout the year, but up until now typically refused to participate in group discussions. In the quotations below, an ellipsis is used to signify a pause, not omitted text. When Diana asked for her opinion, Ellie pointed to the sides of the mast and said, "I think it is two…that here (pointed to left side of the boat) don't have anything like it... this... like this need 'nother one like that. Because this is two there." Watching her point as she spoke makes clear that she was referring to the uneven number of pipe cleaners fastening the mast. We intuit this was significant to her given that this inconsistency could result in an uneven weight distribution and make the boat unstable.

Quickly after the first comment, Ellie pointed to the mast and said, "has lots of feathers – has a little bit like too heavy and this boat will down here (moved hands from one side of the body to the other, while turning them)." She was now alluding to a different feature of the boat: being top-heavy. Additionally, Ellie gestured with her hands how top-heavy objects fall.

She immediately added, "here (pointed to the pipe cleaners holding base and mast together) is a little bit, and don't have a big – and don't have it," referring to the amount of space between the sides of the sponge and the mast. In other words, Ellie highlighted that the mast was off-center, which would also contribute to the instability of the boat. Finally, Ellie returned to her original remark that two pipe cleaners were holding the left side of the mast, while only one was holding the right side, which she expressed while pointing to the pipe cleaners themselves [see video, 12].

Ellie's explanation foregrounds the affordances of a learning environment that engages students in inductive reasoning. First, it is important to highlight that the tangible and shared aspects of students' observations were crucial in generating ideas and conversations about them. Ellie seemed as surprised as her peers by the fact that their predictions had been contradicted by the boat capsizing. She seemed determined to make sense of which features of the boat had contributed to the instability. She listened to her peers' ideas about the sponges, but decided there were other factors that had not been addressed. And while she did not make generalizations from her observations with regards to the relationships between balancing and floating, it is clear that her evidence-based claims were solid first steps towards abstracting principles. These features of the learning environment allowed Ellie to experience productive disciplinary engagement as she spontaneously and substantially contributed to the conversation.

Participating in this type of discussion appears to have bolstered Ellie's confidence to participate despite her limited English vocabulary. Perhaps the sense she had made about the situation compelled her to contribute her views, a compulsion that may have trumped her language inhibition. Either way, the shared experiences and setting led to greater participation from this student. It is difficult to imagine how Ellie's statement could have been as nuanced without the experience and the object she could continuously refer to. We conjecture that students participated in the conversation because the linguistic complexity remained at a level that was comprehensible. In Ellie's particular case, we see how resorting to a range of communicative strategies, like speech and gesturing were effective at getting her point across. These features of the learning environment appear to have attended to the linguistic needs of ELLs and supported the continuous development of language skills.

## CONCLUSIONS AND IMPLICATIONS

Schools are educating more students whose first language is other than English. Therefore, it is important to consider what kinds of learning environments are most effective at engaging these students. In this study, we argued that classrooms based on the principles of scientific induction and evidenced-based reasoning could be very effective at accomplishing this task. Students relied on previous experiences and in-class shared observations to make claims about mechanisms they identified as salient.

This classroom, which promoted scientific induction, successfully supported students in participating in class discussions.

Simultaneously, we claim, this space was also well-suited for supporting ELLs in their path towards bilingualism. The data suggest that an important aspect of these environments is the shared, tangible experience that is the basis of thoughtful reasoning, and expressed in public, collaborative discussions. Activities like the one presented above may be ideal for creating situations in which ELLs can have meaningful and authentic conversations, contributing to the development of language skills.

These findings have significant research and pedagogical implications. First and foremost, we assert physics is a particularly good context for enacting Krashen's model for second language acquisition. Moreover, the data suggests the educational system should not separate language acquisition from other subjects, like physical sciences. Instead, it is important to design learning environments and activities that foster the productive disciplinary engagement of ELLs through evidence-based reasoning.

## ACKNOWLEDGMENTS

We thank Krysta and her students for the opportunity to work alongside and learn from them. We also thank Elizabeth Olivares for bringing Krashen's model to our attention.

## REFERENCES

[1] A. L. Kindler. *Survey of the states' LEP students and available educational programs and services: 2000-2001*, U.S Dept of Education - OELA, Washington DC, 2002.
[2] J. McMurrer. *Instructional Time in Elementary Schools*. Washington, DC, 2008
[3] E. Suarez and V. Otero. *PERC Proceedings 2012* (AIP, Melville, NY) p. 406 (2012)
[4] B. Warren *et al.*, *JRST*, **38**(5), 529-552 (2001)
[5] E. Wenger. *Communities of Practice*. Cambridge University Press, NY, 1998.
[6] C. Baker. *Foundations of Bilingual Education and bilingualism*. Clevedon. Philadelphia, PA, 1993.
[7] R. Engle and F. Conant. *Cognition & Instruction*, **20**(4), 399-483 (2002)
[8] C. R. Mann. *The Teaching of Physics for Purposes of General Education*. MacMillan, NY, 1912.
[9] L. Vygotsky, *Thought and Language*, MIT Press, Cambridge, 1962.
[10] V. Otero and M. Nathan, *JRST* **45**(4), 497-523 (2008).
[11] K. Gutierrez *et al.*, *Lang Arts* **74**(5), 368-378 (1997).
[12] http://bit.ly/11T0YwG (email first author for access).